# Analytically Solvable 2D Potentials: Bound States

S. M. Al-Marzoug[a,b],
H. Bahlouli[a,b] and M. S. Abdelmonem[a,b]

[a] *Physics Department, King Fahd University of Petroleum & Minerals, Dhahran 31261, Saudi Arabia*
[b] *Saudi Center for Theoretical Physics, 31261, Saudi Arabia*

Using a suitable Laguerre basis set that ensures a tridiagonal matrix representation of the reference Hamiltonian, we were able to evaluate in closed form the matrix representation of the associated Hamiltonian for few exactly solvable 2D potentials. This enabled us to treat analytically the full Hamiltonian and compute the associated bound states spectrum as the eigenvalues of the associated analytical matrix representing their Hamiltonians. Finally we compared our results satisfactorily with those obtained using the Gauss quadrature numerical integration approach.

Two-dimensional quantum systems constitutes an intermediate step between the complexity of the three dimensional one and the simplicity of the one dimensional system. Their experimental realization can be achieved in many way, for example, if the motion of the electron around the proton is constrained to be planar (say, by applying a strong magnetic field) then this problem will considered within the context of quantum mechanics as a two-dimensional hydrogen atom. There are many physical applications in which systems are effectively two-dimensional (e.g., adsorbed atoms on surfaces that behave like 2D at low temperatures). In all these studies, an important topological question that is of concern is the role played by the number of spatial dimensions. In addition, two-dimensional space has its own merit and in many circumstances physical results cannot be derived simply from its three dimensional counterpart. For example, Poisson's equation for a point charge in 2D is solved for $\ln r$ rather than the $r^{-1}$ in three dimensions. Evidently, the singularity at the origin in this case is stronger in 2D than in 3D. For further discussion on the relevance of the two-dimensional hydrogen atom and its relevance in physical applications, we refer the reader to Dasgupta's work [1].

One and two-dimensional hydrogen atoms were studied using space reduction from both Cartesian and spherical coordinate approach [2]. Through such an exercise, one might gain insight into the electronic structure in a two-dimensional space and could address the consequences of losing one degree of freedom. When studying the two-dimensional hydrogen atom it was found that, aside from the shift in the electron cloud towards the nucleus, the radial wave function is very similar to that in the 3D case; both are written in terms of the associated Laguerre polynomials [3]. In fact, we will see that the 2D results will be closely related to those in 3D and that the cylindrical Bessel functions in 2D will be replaced by spherical Bessel functions in 3D. Scattering in 2D was studied in the past by Lapidus [4] who presented a partial wave analysis of the problem. Maurone [5] derived the analogous expression for the optical theorem in two dimensions.

The Schrödinger equation for most of the 2D potentials could not be solved analytically; hence various numerical and perturbative methods have been devised in order to obtain the energy levels and related physical quantities. Even for short-range potentials, the $r^{-2}$ behavior due to the centrifugal term makes the task of obtaining accurate numerical solutions a non-trivial task. Our approach constitutes a significant contribution in this regard. It employs the unique feature of our Laguerre basis set which allows us to evaluate analytically all matrix elements of the Hamiltonians associated with some specific potentials which are then said to be analytically solvable according to our prescription.

Our approach for the study of analytically solvable potentials (for any angular momentum) is inspired by the J-matrix method [6], an algebraic method for extracting resonance and bound states information using computational tools devised entirely in square integrable bases. In this approach, the total Hamiltonian is written as a sum two parts: a reference Hamiltonian $H_0$ which is treated exactly and analytically and the remaining part which is treated numerically. The discrete $L^2$ bases used in the calculation and analysis are required to carry a tridiagonal matrix representation for the reference wave operator. Moreover, the use of discrete basis sets offers considerable advantage in the calculation of bound states and resonances because it is an algebraic scheme that requires only standard matrix technique rather than the usual approach of numerical integration of the differential equation.

The two-dimensional time-independent Schrödinger wave equation for a point particle of mass $m$ in a spherically symmetric potential $V(r)$ reads as follows

$$(H-E)\psi(r,\phi) = \left[-\frac{\hbar^2}{2m}\left(\frac{1}{r}\frac{\partial}{\partial r}\left(r\frac{\partial}{\partial r}\right) + \frac{1}{r^2}\frac{\partial^2}{\partial \phi^2}\right) + V(r) - E\right]\psi(r,\phi) = 0, \qquad (1)$$

In the J-matrix approach, the wave function $\psi$ is expanded in the space of square integrable functions with discrete basis elements $\{\phi_n\}_{n=0}^{\infty}$ as $|\psi(\vec{r},E)\rangle = \sum_n f_n(E)|\phi_n(\vec{r})\rangle$, where $\vec{r}$ is the set of coordinates for real space and $E$ is the system's energy. The basis functions must be compatible with the domain of the Hamiltonian and satisfy the vanishing boundary conditions at $r = 0$ and $r = \infty$. Equation (1) can be separated owing to the spherical symmetry of the potential as $\psi(r,\phi) = R(r)e^{i\ell\phi}$, with $\ell = 0, \pm1, \pm2,\ldots$etc. Then letting $R(r) = \chi(r)/\sqrt{r}$ we obtain the effective radial equation

$$(H-E)\chi(r) = \left[-\frac{1}{2}\frac{d^2}{dr^2} + \frac{\ell^2 - 1/4}{2r^2} + V(r) - E\right]\chi(r) = 0 \qquad ;\ell = 0, \pm1, \pm2, \pm3\cdots \quad (2)$$

We have used the atomic units $\hbar = m = 1$ where length is measured in units of $a_0 = 4\pi\epsilon_0\hbar^2/m$ (for an electron, this is the Bohr radius) and m is the electron mass. In this work we will choose a complete $L^2$ basis set $\{\phi_n\}$ to make the matrix representation of the reference Hamiltonian tridiagonal. The following choice of basis functions [7], called the Laguerre basis, is compatible with the domain of the Hamiltonian, satisfies the desired boundary conditions, and will results in a tridiagonal matrix representation for $H_0$.

$$\phi_n(x) = a_n x^\alpha e^{-x/2} L_n^\nu(x) \quad ; \quad x = \lambda r \tag{3}$$

where $\lambda$ is a positive length scale parameter, which allows for more computational freedom, $\alpha > 0$ and $\nu$ are basis parameters to be selected so as to make the matrix representation of $H_0$ tridiagonal. $L_n^\nu(x)$ is the Laguerre polynomial of degree $n$ and $a_n$ is the normalization constant $\sqrt{\lambda \Gamma(n+1)/\Gamma(n+\nu+1)}$. Now, the only remaining quantity that is needed to perform the calculation is the matrix elements of the effective potential $V(r)$. This is obtained by evaluating the integral

$$V_{nm} = \int_0^\infty \phi_n(\lambda r) V(r) \phi_m(\lambda r) dr = \lambda^{-1} a_n a_m \int_0^\infty x^{2\alpha} e^{-x} L_n^\nu(x) L_m^\nu(x) V(x/\lambda) dx. \tag{4}$$

The evaluation of such an integral for a general effective potential is almost always done numerically using the Gauss quadrature approximation [8]. However, in our present work we are interested in situations where this matrix element can be either evaluated analytically or can be put in a tridiagonal form along with $H_0$.

To proceed further in our computations we need to select the solvable potentials we want to treat using our tridiagonal procedure. First we consider the famous Yukawa potential also called the screened Coulomb potential used in various areas of physics to model singular but short-range interactions [9]. In our present work, however, we consider a generalized form for this potential described by

$$V(r) = -\frac{A}{r} e^{-\mu r} \quad ; \quad \mu = \mu_R + i \mu_I, \tag{5}$$

where $A$, $\mu_R$ and $\mu_I$ are real positive parameters describing the strength of the potential and the real and imaginary parts of the screening parameter. The advantage of the general formulation is that it gives the classical Yukawa potential from (5) by choosing a real screening parameter, $\mu_I = 0$, while the cosine like Yukawa potential and the sine-like Yukawa potential are obtained by taking the real and imaginary parts of (5), respectively. In the case of Yukawa potential our $H_0$ is selected to be

$$H_0 = -\frac{1}{2} \frac{d^2}{dr^2} + \frac{\ell^2 - 1/4}{2r^2} \tag{6}$$

Using the standard J-matrix manipulation, we obtain the following tridiagonal matrix representation for $H$ [10]

$$\tfrac{8}{\lambda^2}(H_0)_{nm} = (2n+\nu+1)\delta_{n,m} + \sqrt{n(n+\nu)}\delta_{n,m+1} + \sqrt{(n+1)(n+\nu+1)}\delta_{n,m-1}, \tag{7}$$

by selecting our parameter so that $\nu = 2\ell$ and $\alpha = \ell + \frac{1}{2}$. In the manipulation, we used the differential equation, differential formula, three-term recursion relation, and orthogonality formula of the Laguerre polynomials [11]. The basis $\{\phi_n\}$ is not orthogonal but trithogonal. That is, its overlap matrix

$$\langle \phi_n | \phi_m \rangle = (2n+\nu+1)\delta_{n,m} - \sqrt{n(n+\nu)}\delta_{n,m+1} - \sqrt{(n+1)(n+\nu+1)}\delta_{n,m-1}, \tag{8}$$

is tridiagonal. Now, the only remaining quantity that is needed to perform the calculation is the matrix elements of the potential $V(r)$ given by (4), more explicitly

$$V_{nm} = \frac{1}{\lambda} \int_0^\infty \phi_n(x) V(x/\lambda) \phi_m(x) dx = -A a_n a_m \int_0^\infty x^\nu e^{-\sigma x} L_n^\nu(x) L_m^\nu(x) dx \tag{9}$$

where $x = \lambda r$ and $\sigma = 1 + \mu/\lambda$. Using the following integral result [12]

$$\int_0^\infty x^\nu e^{-\sigma x} L_n^\nu(x) L_m^\nu(x) dx$$
$$= \frac{\Gamma(n+m+\nu+1)}{\Gamma(n+1)\Gamma(m+1)} \frac{(\sigma-1)^{m+n}}{\sigma^{m+n+\nu+1}} {}_2F_1\left(-n,-m,-n-m-\nu;\frac{\sigma(\sigma-2)}{(\sigma-1)^2}\right) \quad (10)$$

the matrix elements of the Yukawa potential become

$$V_{nm} = -A a_n a_m \frac{\Gamma(n+m+\nu+1)}{\Gamma(n+1)\Gamma(m+1)} \frac{(\sigma-1)^{m+n}}{\sigma^{m+n+\nu+1}} {}_2F_1\left(-n,-m,-n-m-\nu;\frac{\sigma(\sigma-2)}{(\sigma-1)^2}\right). \quad (11)$$

Now we fix the basis dimension N and evaluate the matrix representation of our full Hamiltonian whose eigenvalues will represent the bound states of our system. For the sake of simplicity we just consider the cosine-Yukawa potential which will then be obtained if we take the real part of the matrix element (11). For simplicity we limit our computations to the special case $\mu_R = \mu_I = \delta$ and consider only the cosine-like Yukawa potential (recall that our present formalism allows for cosine-like and sine-like Yukawa in addition to the usual non oscillating Yukawa potential, all of them treated on equal footing).

Our calculation strategy is as follows. For a given choice of physical parameters, we investigate the stability of calculated eigenvalues that correspond to bound states as we vary the scaling parameter $\lambda$ until we reach a plateau in $\lambda$ [13]. Then to improve on the accuracy of the results, we selected a value of $\lambda$ from within the plateau and increase the size $N$ of the Hamiltonian matrix until the desired accuracy is reached. Our calculations show that the stability plateau for numerical computations becomes narrower as we get closer and closer to the critical value, $\delta_c$, associated with bound-unbound transition.

In Table 1, we show the bound state energies for the s-states $(\ell = 0)$ with quantum number n = 1, 2 and 3 for different values of $\delta$, and compare our results with Gauss quadrature approach. The parameters used are $A = 1$, $N = 100$ and $\lambda$ in the stability range from 1 to 5. As reflected in this table our results using analytical approach are in good agreement with those obtained from Gauss quadrature approach. The accuracy of our results reduces as we get close to the critical value of the screening parameter, $\delta_c(n\text{s})$, defined to be the value of the screening parameter at which the ns bound state disappears and emerges as a resonance. In general, the number of significant figures for values of the screening parameter away from the critical value are large because being away from $\delta_c$ the wave function is very much localized and hence can be described by few elements in the basis set to reach the desired accuracy. On the contrary for values close to $\delta_c$ the wave function start having a long range tail and under these circumstances, the most suitable basis set should have long extensions (small $\lambda$) and/or a bigger size (large $N$) to ensure that the potential is sampled correctly in regions away from the origin.

Next we consider a more general situation where we look for a potentially solvable potential in our tridiagonal representation in the general Laguerre basis set defined by (3) and compute the matrix element of each element in (2) which give after a lengthy algebra to ( we define J = H – E )

$$\frac{2}{\lambda^2} J_{m,n} = \frac{2}{\lambda^2} \langle \phi_m | (H-E) | \phi_n \rangle = \frac{a_n}{a_{n-1}} (n+\nu)(2\alpha - \nu - 1) \langle \phi_m | \frac{1}{y} | \phi_{n-1} \rangle$$
$$+ \langle \phi_m | \frac{n}{y}\left(1 + \frac{\nu + 1 - 2\alpha}{y}\right) + \frac{\ell^2 - \frac{1}{4}\alpha(\alpha - 1)}{y^2} + \frac{\alpha}{y} - \frac{1}{4} + \frac{2}{\lambda^2}(V-E) | \phi_n \rangle \quad (12)$$

If we look for possible forms of potential *V(y)* so as to give rise a tridiagonal representation of the above matrix element, using the orthogonality relation of the Laguerre polynomials, we then end up with two possibilities one for scattering states ( E > 0 ) and one for bound states ( E < 0 ). The bound state situation obtains for the following set of parameters and potential solution

$$\nu = 2\alpha - 2 \quad ; \quad V(r) = \frac{A}{r} + \frac{B}{2r^2} \quad (13)$$

where A and B are potential parameters. This solvable potential is known under the name of Kratzer molecular potential, it has been used extensively to describe molecular systems [14]. Substituting these values in our previous matrix element (12) we obtain

$$\frac{2}{\lambda^2} J_{m,n} = [(2n+\nu+1)(\frac{1}{4} - \frac{2E}{\lambda^2}) + 2\frac{A}{\lambda}]\delta_{mn}$$
$$- (\frac{1}{4} + \frac{2E}{\lambda^2})\left[\sqrt{n(n+\nu)}\delta_{m,n-1} - \sqrt{(n+1)(n+\nu+1)}\delta_{m,n+1}\right] \quad (14)$$

The associated bound states can be obtained by requiring a diagonal representation of (14) and hence results in the following requirements

$$(2n+\nu+1)(\frac{1}{4} - \frac{2E}{\lambda^2}) + 2\frac{A}{\lambda} \quad ; \quad \frac{1}{4} + \frac{2E}{\lambda^2} = 0 \quad (15)$$

which leads to the following closed form for the bound state energies associated with the Kratzer potential given by (13)

$$E = -\frac{1}{2}\left(\frac{4A}{2n+\nu+1}\right)^2 = -\frac{A^2}{2}\left(n + \frac{1}{2} + \sqrt{B + \ell^2}\right)^{-2} \quad (16)$$

Using table of integrals we can write the following closed form expressions for the potential components
$V_{nm} = V_{nm}^{(1)} + V_{nm}^{(2)}$; where

$$V_{nm}^{(1)} = A a_n a_m \begin{pmatrix} \frac{\Gamma(\nu+n+1)}{n!} & ; m=n \\ 0 & ; m \neq n \end{pmatrix},$$

$$V_{nm}^{(2)} = \frac{\lambda B}{2} a_n a_m \Gamma(\nu) \sum_{k=0}^{\min(n,m)} \binom{m+\nu}{m-k}\binom{n+\nu}{n-k}\binom{k+\nu-1}{k} \times {}_2F_1(k-m,\nu+k+1;k+\nu+1;1) \, {}_2F_1(k-n,\nu+k;k+\nu+1;1) \quad (17)$$

Where we denote the first term by $V^{(1)}$ and the second one by $V^{(2)}$

In Table 2 we show the numerical computations of the bound states energies for the Kratzer potential as obtained from the exact formulae (14, 17). In this table we show the first five computed bound states ( energy eigenvalues, -E) (All in atomic units) of the Kratzer potential for A=1 and different values of A and $\ell$. Our computational parameters were N=100 and $\lambda$ within the stability range between 0.3 and 3. Comparison between our results for the bound states and those generated Gauss quadrature suggest that our analytical method is very satisfactory.

Last, but not the least we consider the generalized Morse potential. The Morse potential plays a dominant role in describing the interaction of atoms in diatomic and even polyatomic molecules [15-17]. The effective potential in this case is the sum of the centrifugal potential term that depends on the orbital angular quantum number $\ell$ and the Morse potential. The bound states of this system can be obtained only through numerical approaches where several approximation techniques have been proposed and extensively used in three dimensions with varying degrees of accuracy and stability [18]. We define the Morse potential by

$$V(r) = V_0 ( e^{-2\alpha(\frac{r}{r_0}-1)} - 2\beta e^{-\alpha(\frac{r}{r_0}-1)} ) \qquad (18)$$

Where $V_0$ is the dissociation energy, $r_0$ is the equilibrium inter-nuclear distance, and α is a parameter that controls the width of the potential well and β is an additional potential parameter usually set to unity in the classical Morse potential. In this case the reference Hamiltonian, $H_0$, which is exactly solvable contain only the kinetic and centrifugal terms and is given by (6) while its matrix element is given by (7) and the overlap integral by (8). Using table of integrals and writing the potential as $V(r) = V^1(r) + V^2(r)$, we can write the following closed form expressions for the potential components :

$$V^1_{nm} = \frac{V_0 a_n a_m}{\lambda} e^{2\alpha} \frac{\Gamma(\nu+2)}{\sigma_1^{\nu+2}} \sum_{k=0}^{\min(n,m)} \binom{m+\nu}{m-k}\binom{n+\nu}{n-k}\binom{k+\nu+1}{k} \sigma_1^{-2k} \times$$

$$\times {}_2F_1\left(k-m, \nu+k+2; k+\nu+1; \frac{1}{\sigma_1}\right) {}_2F_1\left(k-n, \nu+k+2; k+\nu+1; \frac{1}{\sigma_1}\right)$$

$$V^2_{nm} = \frac{2\beta V_0 a_n a_m}{\lambda} e^{\alpha} \frac{\Gamma(\nu+2)}{\sigma_2^{\nu+2}} \sum_{k=0}^{\min(n,m)} \binom{m+\nu}{m-k}\binom{n+\nu}{n-k}\binom{k+\nu+1}{k} \sigma_2^{-2k} \times \qquad (19)$$

$$\times {}_2F_1\left(k-m, \nu+k+2; k+\nu+1; \frac{1}{\sigma_2}\right) {}_2F_1\left(k-n, \nu+k+2; k+\nu+1; \frac{1}{\sigma_2}\right)$$

$$\sigma_1 = 1 + \frac{2\alpha}{\lambda r_0}; \sigma_2 = 1 + \frac{\alpha}{\lambda r_0}$$

Thus the Morse potential is not tridiagonal in this basis set but we can still perform easily the computations of the bound states by computing the eigenvalues of the associated Hamiltonian matrix. In Table 3 we present our results for the bound states associated with the generalized Morse potential for different values of the potential parameters for matrix of size N =70 and basis parameter λ within the stability range between 10 and 15. We see from this table that these eigenvalues are in excellent agreement with those computed using Gauss quadrature approach up to 12 digits for low-level excitations. In the table, we had to go to higher order bound states to reach a level at which our results start deviating measurably from the Gauss quadrature one. It is worth mentioning that significant increase in the accuracy can still be achieved by moderately enlarging the basis size *N* even, say, up to *N* = 100. It is also worth noting that the final choice of the parameter *λ* for a given molecule and basis is made only after studying the effect of its variation on the whole energy spectrum

In conclusion, the proposed analytical approach give a very compact closed form for the potential matrix elements in a suitable basis set and hence enhances the accuracy of computations of the bound states energies by allowing a complete analytic treatment of the full Hamiltonian matrix elements. The desired numerical precision, in our case, is limited only by the size of the basis set and machine accuracy. As explained in the introduction the real power of our approach is that: (1) it allows for the analytic computation of the matrix elements of the reference Hamiltonian, and (2) it provides for an analytical expression of the potential matrix elements in a suitable selected basis set. The eigenvalue computation of the resulting full Hamiltonian is performed to the desired accuracy using two free parameters, the scaling length of the basis set parameter $\lambda$ and the dimension of the basis space $N$. To illustrate the accuracy of our approach, we used it to calculate bound states energies for the selected potential that admit a analytical closed formula for the associated potential matrix elements. Comparison between bound states generated by our analytical approach and those from numerical computations, using Gauss quadrature approach, suggest that our analytical approach is efficient and reliable.

## ACKNOWLEDGMENTS

The authors acknowledge the support provided by the Physics department at King Fahd University of Petroleum & Minerals under research grant RG1109-1-2 and the Saudi Center for Theoretical Physics (SCTP).

**REFERENCES:**
[1]  B.B. Dasgupta, Am. J. Phys. 49, 189 ( 1981)
[2]  G. Q. Hassoun, Am. J. Phys. 49, 143 (1981).
[3]  B. Zaslow and M. E. Zandler, Am. J. Phys. 35, 1118 ( 1967)].
[4]  I. R. Lapidus, Am. J. Phys. 50, 45 (1982).
[5]  P.A. Maurone, Am. J. Phys. 51, 856 (1983).
[6]  E. J. Heller and H. A. Yamani, Phys. Rev. A **9**, 1201 (1974); *ibid.* 1209 (1974); A. D. Alhaidari, E. J. Heller, H. A. Yamani, and M. S. Abdelmonem (Eds.), *The J-matrix method: Recent Developments and Selected Applications* (Springer, Heidelberg, 2008).
[7]  H. A. Yamani and L. Fishman, J. Math. Phys. **16**, 410 (1975);
     A. D. Alhaidari, Ann. Phys. **317**, 152 (2005).
[8]  See, for example, Appendix A in: A. D. Alhaidari, H. A. Yamani, and M. S. Abdelmonem, Phys. Rev. A **63**, 062708 (2001).
[9]  H. Yukawa, Proc. Phys. Math. Soc. Jpn. **17**, 48 (1935).
[10] A. D. Alhaidari, International Journal of Modern Physics A Vol. 20, No. 12, 2657 (2005).
[11] W. Magnus, F. Oberhettinger, and R. P. Soni, *Formulas and Theorems for the Special Functions of Mathematical Physics*, 3$^{rd}$ edn. (Springer-Verlag, New York, 1966) pp. 239-249; M. Abramowitz and I. A. Stegun (eds.), *Handbook of Mathematical Functions* (Dover, New York, 1964).
[12]  I. S. Gradshteyn and I. M. Ryzhik, " Table of integrals, series, and products", Academic Press, 7$^{th}$ edition (2007) [ formula 7.414 (4) page 809 ].
[13] I. Nasser, M. S. Abdelmonem, H. Bahlouli and A. D. Alhaidari, *J. Phys. B: At. Mol. Opt. Phys.* **40** (2007) 4245.
[14]  R. J. Le Roy, R. B. Bernstein 1970 *J Chem Phys*, **52** 3869;
      O. Bayrak, I. Boztosun, H. Ciftci, 2007 *Int. J. Quant. Chem.* **107** 540;
      M. Sameer, M. Ikhdair and S. Ramazan 2009 *J. Math. Chem.* **45** 1137,
      and references there in.
[15]   P.M. Morse, Phys. Rev. 34 (1929) 57.
[16]   S.H. Dong, R. Lemus, A. Frank, Int. J. Quantum Phys. 86 (2002) 433, and references therein.
[17]   S. Flügge, Practical Quantum Mechanics, vol. I, Springer-Verlag, Berlin, 1994.

[18]  C.L. Pekeris, Phys. Rev. 45 (1934) 98; R. Herman, R.J. Rubin, Astrophys. J. 121 (1955) 533; M. Duff, H. Rabitz, Chem. Phys. Lett. 53 (1978) 152; J.R. Elsum, G. Gordon, J. Chem. Phys. 76 (1982) 5452; E.D. Filho, R.M. Ricotta, Phys. Lett. A 269 (2000) 269; D.A. Morales, Chem. Phys. Lett. 394 (2004) 68;  M. Bag, M.M. Panja, R. Dutt and Y. P. Varshni, Phys. Rev. A 46 (1992) 6059; C. Berkdemir, Nucl. Phys. A 770 (2006) 32; O. Bayrak, I. Boztosum, J. Phys. A: Math. Gen. 39 (2006) 6955; E. Castro, J. L. Paz and P. Martin, Journal of Molecular Structure: THEOCHEM 769, 15 ( 2006).

**TABLE CAPTIONS:**

**Table 1:** S-wave bound state energies (-E) for the cosine-Yukawa potential with parameters A = 1, λ in the stability range from 1 to 5, N = 100 and for different values of the screening parameter.

**Table 2:** The first five computed eigenvalues (-E) (All in atomic units) of the Kratzer potential for A=1 and different values of B and $\ell$. Our parameters are N=100. λ within the range between 0.3 and 3

**Table 3**: Bound state energy eigenvalues (-E) (All in atomic units) of the generalized Morse potential for different values of parameters. Our parameters are N =70. λ within the range between 10 and 15

**Table 1**

| δ | Analytical approach | Gauss Quadrature approach |
|---|---|---|
| 0.01 | 1.9900001243765<br>0.2122269805218<br>0.07003239211313<br>0.03093144647664<br>0.0149826725629<br>0.0071278983166<br>0.0029082843879<br>0.0006477201 | 1.9900001243765<br>0.2122269805218<br>0.07003239211313<br>0.03093144647664<br>0.01498267256299<br>0.00712789831668<br>0.00290828438798<br>0.0006477201 |
| 0.08 | 1.9200614950617<br>0.1442942629202<br>0.0118175986105 | 1.9200614950617<br>0.1442942629202<br>0.0118175986105 |
| 0.1 | 1.9001189204077<br>0.1261021700846<br>0.00187488962075 | 1.900118920407<br>0.126102170084<br>0.00187488962075 |
| 0.2 | 1.8009057424238<br>0.048234960913 | 1.8009057424238<br>0.0482349609138 |
| 0.5 | 1.5123062833952 | 1.51230628339522 |
| 1 | 1.08022847887960 | 1.08022847887961 |
| 2 | 0.458673666401 | 0.458673666401 |
| 5 | 0.0087175321 | 0.008717532 |
| 9 | 8.6595*E-6 | 8.6593*E-6 |

**Table 2**

| B | n | $\ell = 1$ | | | $\ell = 2$ | | | $\ell = 5$ | | |
|---|---|---|---|---|---|---|---|---|---|---|
| | | Exact | Analytical approach | Gauss Quadrature approach | Exact | Analytical approach | Gauss Quadrature approach | Exact | Analytical approach | Gauss Quadrature approach |
| 50 | 0 | 0.008562900642375 | 0.0085629006423 | 0.0085629006423 | 0.008117084827976 | 0.008117084827 | 0.008117084827 | 0.005958747303690 | 0.0059587473036 | 0.0059587473036 |
| | 1 | 0.006695745370544 | 0.0066957453705 | 0.0066957453705 | 0.006386070585535 | 0.006386070585 | 0.006386070585 | 0.004843517472534 | 0.0048435174725 | 0.0048435174725 |
| | 2 | 0.005378822847548 | 0.0053788228475 | 0.0053788228475 | 0.005155045938050 | 0.005155045938 | 0.005155045938 | 0.004014411087421 | 0.0040144110874 | 0.0040144110874 |
| | 3 | 0.004415400957402 | 0.0044154009574 | 0.0044154009574 | 0.004248475141182 | 0.004248475141 | 0.004248475141 | 0.003381307818617 | 0.0033813078186 | 0.0033813078186 |
| | 4 | 0.003689414577626 | 0.0036894145776 | 0.0036894145776 | 0.003561603048495 | 0.003561603048 | 0.003561603048 | 0.002886964601972 | 0.0028869646019 | 0.0028869646019 |
| 5 | 0 | 0.057474635269819 | 0.0574746351 | 0.0574746354 | 0.040816326530612 | 0.040816326530612 | 0.040816326530612 | 0.013994929411735 | 0.0139949294117 | 0.0139949294117 |
| | 1 | 0.032054427436461 | 0.0320544273 | 0.0320544275 | 0.024691358024691 | 0.024691358024691 | 0.024691358024691 | 0.010270804820936 | 0.0102708048209 | 0.0102708048209 |
| | 2 | 0.020410288672876 | 0.0204102886 | 0.0204102887 | 0.016528925619834 | 0.016528925619834 | 0.016528925619834 | 0.007857171966530 | 0.0078571719665 | 0.0078571719665 |
| | 3 | 0.014125719046627 | 0.0141257190 | 0.0141257190 | 0.011834319526627 | 0.011834319526627 | 0.011834319526627 | 0.006204199126583 | 0.0062041991265 | 0.0062041991265 |
| | 4 | 0.010352951221794 | 0.0103529512 | 0.0103529512 | 0.008888888888888 | 0.008888888888888 | 0.008888888888888 | 0.005022852463037 | 0.0050228524630 | 0.0050228524630 |
| 1 | 0 | 0.136454928592147 | 0.136453616 | 0.136461512 | 0.066790737340724 | 0.066790735906343 | 0.0667907582895 | 0.015949462144524 | 0.0159494621445 | 0.0159494621445 |
| | 1 | 0.058874503045718 | 0.058874079 | 0.058876679 | 0.035821227603347 | 0.035821226852212 | 0.0358212387831 | 0.011481831764407 | 0.0114818317644 | 0.0114818317644 |
| | 2 | 0.032634801055626 | 0.032634619 | 0.032635742 | 0.022291236000336 | 0.022291235591735 | 0.0222912421493 | 0.008658743703939 | 0.0086587436962 | 0.0086587436962 |
| | 3 | 0.020704366750212 | 0.020704273 | 0.020704852 | 0.015196424803818 | 0.015196424562388 | 0.0151964284637 | 0.006761952806216 | 0.0067619299812 | 0.0067619299862 |
| | 4 | 0.014294731377207 | 0.014294677 | 0.014295012 | 0.011019378022041 | 0.011019377868956 | 0.0110193803550 | 0.005426455616861 | 0.0054216132940 | 0.0054216140253 |
| 0.1 | 0 | 0.208436783273251 | 0.208436756 | 0.208437849 | 0.078433271272334 | 0.078433271262322 | 0.07843327268673 | 0.016469043621932 | 0.0164690436219 | 0.0164690436219 |
| | 1 | 0.076965389599071 | 0.076965383 | 0.076965650 | 0.040242948225715 | 0.040242948220950 | 0.04024294890538 | 0.011798026267925 | 0.0117980262679 | 0.0117980262679 |
| | 2 | 0.039701305667814 | 0.039701302 | 0.039701403 | 0.024420944763840 | 0.024420944761374 | 0.02442094511732 | 0.008865256070540 | 0.0088652560668 | 0.0088652560668 |
| | 3 | 0.024164322881252 | 0.024159070 | 0.024159128 | 0.016380596093781 | 0.016380596092368 | 0.01638059629706 | 0.006904176778903 | 0.0069041631880 | 0.0069041631883 |
| | 4 | 0.016239418588263 | 0.015703849 | 0.015703850 | 0.011744364350880 | 0.011744364350019 | 0.01174436447747 | 0.005528532691013 | 0.0055250410995 | 0.0055250411548 |

**Table 3**

| $\ell$ | $r_0$ | $\alpha$ | $V_0$ | β=0.8 | | β=1 | | β=1.2 | |
|---|---|---|---|---|---|---|---|---|---|
| | | | | Analytical approach | Gauss Quadrature approach | Analytical approach | Gauss Quadrature approach | Analytical approach | Gauss Quadrature approach |
| 0 | 1 | 2 | -10 | 241.4455469169<br>92.2621055918<br>21.147768355290 | 241.4455469169<br>92.26210559186<br>21.14776835529 | 216.47559486094<br>73.0655914016<br>7.617263800989 | 216.47559486093<br>73.06559140163<br>7.6172638009 | 191.571748471411<br>54.10423557665 | 191.5717484714<br>54.1042355766 |
| 2 | 1 | 2 | -10 | 73.8752316465212<br>11.7565360884 | 73.87523164652<br>11.75653608845 | 55.151473935195 | 55.151473935195 | 36.687751877201 | 36.687751877201 |
| 0 | 4 | 1.5 | -6 | 55.042767263132<br>33.964364044587<br>20.62526735064<br>11.160124151925<br>4.32512276547 | 55.04276726313<br>33.964364044588<br>20.62526735064<br>11.160124151925<br>4.32512276547 | 45.203869139509<br>25.21574484179<br>12.801414587566<br>4.23015943520 | 45.2038691395<br>25.21574484179<br>12.80141458756<br>4.230159435 | 35.38643997352<br>16.526155682649<br>5.087723194206 | 35.386439973520<br>16.52615568264<br>5.08772319420 |
| 1 | 4 | 1.5 | -6 | 41.8433860514363<br>26.11877926631<br>15.1603025679012<br>7.222263081408<br>1.53509424176 | 41.843386051436<br>26.118779266314<br>15.160302567901<br>7.222263081408<br>1.535094241764 | 32.629596329074<br>17.865063073765<br>7.80093221213<br>0.7682013249054 | 32.629596329074<br>17.86506307376<br>7.80093221213<br>0.768201324905 | 23.456123825950<br>9.69454254105<br>0.58960423692 | 23.456123825950<br>9.694542541058<br>0.589604236925 |
| 2 | 4 | 1.5 | -6 | 31.80730946194<br>19.317248416896<br>10.269437452762<br>3.691373989975 | 31.807309461944<br>19.317248416896<br>10.26943745276<br>3.691373989975 | 23.136414765140<br>11.549044910706<br>3.389515557097 | 23.136414765140<br>11.5490449107062<br>3.389515557097 | 14.526564734920<br>3.894631587677 | 14.52656473492<br>3.89463158767787 |